\documentclass[a4paper,11pt]{article}
\usepackage{jinstpub} % for details on the use of the package, please see the JINST-author-manual
\usepackage{lineno}
\usepackage{siunitx}
\usepackage{glossaries}
\usepackage{graphicx}
\usepackage[version=4]{mhchem}
\DeclareSIUnit{\u}{u}
\DeclareSIUnit{\sample}{Sa}
\DeclareSIUnit{\electron}{e}

\newacronym{MC}{MC}{Monte Carlo}
\newacronym{FWHM}{FWHM}{full width at half maximum}
\newacronym{LET}{LET}{linear energy transfer}
\newacronym{PHA}{PHA}{pulse height analysis}
\newacronym{CIRT}{CIRT}{carbon-ion radiation therapy}
\newacronym{RBE}{RBE}{relative biological effectiveness}
\newacronym{QA}{QA}{quality assurance}
\newacronym{CSA}{CSA}{charge-sensitive amplifier}
\newacronym{ADC}{ADC}{analog-to-digital converter}
\newacronym{SNR}{SNR}{signal-to-noise ratio}
\newacronym{TEPC}{TEPC}{tissue-equivalent proportional counter}
\newacronym{MCA}{MCA}{multichannel analyzer}
\newacronym{PCB}{PCB}{printed circuit board}
\newacronym{ASIC}{ASIC}{application-specific integrated circuit}
\newacronym{ENC}{ENC}{equivalent noise charge}
\newacronym{PUR}{PUR}{pileup rejection}
\newacronym{FPGA}{FPGA}{field programmable gate array}
\newacronym{DAQ}{DAQ}{data acquisition}
\newacronym{WET}{WET}{water equivalent thickness}
\newacronym{SSD}{SSD}{solid-state detector}
\newacronym{SV}{SV}{sensitive volume}
\newacronym{CVD}{CVD}{chemical vapor deposition}
\newacronym{SiC}{SiC}{silicon carbide}
\newacronym{FIR}{FIR}{finite impulse response}
\newacronym{SoC}{SoC}{system-on-chip}
\newacronym{DMA}{DMA}{direct memory access}
\newacronym{SoM}{SoM}{system-on-module}
\newacronym{DSP}{DSP}{digital signal processing}

\proceeding{27$^{\text{th}}$ International Workshop on Radiation Imaging Detectors (IWORID)\\
June 28, 2026 to July 2, 2026\\
Ghent, Belgium}

\title{Spectacular -- A Modular DAQ System for Microdosimetry}

\author[a,1]{Matthias Knopf, \note{Corresponding author.}}

\author[b]{Simon Waid,}
\author[b,a]{Daniel Radmanovac,}
\author[b,a]{Sebastian Onder,}
\author[b]{Richard Thalmeier,}
\author[b,a]{Peter Fischer,}
\author[b]{Stefan Gundacker,}

\author[c]{Claudio Verona,}
\author[c]{Edoardo Domenicone,}

\author[b]{Thomas Bergauer,}
\author[a]{Albert Hirtl}

\affiliation[a]{TU Wien, Atominstitut,\\Stadionallee 2, 1020 Wien, Austria}
\affiliation[b]{Marietta Blau Institute for Particle Physics (MBI), Austrian Academy of Sciences,\\Dominikanerbastei 16, 1010 Vienna, Austria}
\affiliation[c]{Dip. di Ingegneria Industriale Università di Roma "Tor Vergata", INFN-Roma2, Rome 00133, Italy}

\emailAdd{matthias.knopf@tuwien.ac.at}

\abstract{Particle therapy using light ions like protons, helium-ions or carbon-ions enables precise tumor targeting with enhanced biological effectiveness while minimizing damage to healthy tissue. Successful treatment planning depends not only on the absorbed dose but also on quality of the radiation, as quantified by the linear energy transfer (LET). Microdosimetry provides a direct experimental determination of such quantities by measuring the energy deposited per incoming particle in micrometer-sized solid-state detectors representing the relevant biological scales. However, the small signal amplitudes and high particle rates in therapeutic ion beams (up to \SI{e10}{\per\second}) challenge existing readout systems, which are not sufficiently optimized for reliable operation with respect to pileup and signal-to-noise ratio (SNR). To address this, a modular data acquisition (DAQ) system was developed to accelerate the design of custom readout electronics and sensor characterization in microdosimetry and related spectroscopic applications. Centered around a Xilinx Zynq system-on-chip, it combines real-time processing, high-bandwidth streaming, and high-resolution digitization (16 bit at \SI{100}{\mega\sample\per\second}) to enable advanced digital signal processing. The platform integrates programmable power supplies, a bias-voltage filter, test-pulse generators, and flexible I/O. Detector and preamplifier front-ends are hosted on interchangeable \color{black}daughter boards \color{black} connected via a standardized interface, allowing different hardware configurations and readout algorithms to be evaluated on the same platform. The \emph{Spectacular} DAQ system was successfully tested at the MedAustron ion therapy facility with custom charge-sensitive amplifiers and a diamond microdosimeter. \color{black}First results demonstrate the feasibility and provide a proof-of-concept for potential future applications in clinical practice. \color{black}}


\begin{document}
\maketitle
\flushbottom

\section{Introduction}
\label{sec:intro}

Light-ion therapy is an established cancer treatment modality, increasingly available worldwide as an alternative to conventional photon radiotherapy. Its clinical advantage derives from the favorable depth-dose relation of charged particles, characterized by the Bragg peak, and their enhanced biological effectiveness. Clinical studies in \gls{CIRT} have linked radiation quality to tumor control, mainly using dose-averaged \gls{LET}, $\overline{{LET}}_D$~\cite{Hagiwara_2020,Matsumoto_2020_Chondrosarcoma,Molinelli_2021_Sacral_Chordoma}. Furthermore, recent cell irradiation studies with protons show that similar $\overline{{LET}}_D$ values can lead to different biological responses when beam compositions differ, suggesting that the entire distribution may provide relevant information beyond average values~\cite{Guan_2024_Cope_for_uDos}. This motivates the introduction of radiation quality metrics into treatment planning and clinical workflows~\cite{Kalholm_2021_LET_Review,Magini_2025_LUT}. As \gls{LET} is not directly measurable and is commonly obtained through \gls{MC} simulations, microdosimetry offers an experimental approach for characterizing radiation quality. Its clinical implementation may support both \gls{RBE}-based treatment optimization and \gls{QA} applications, including independent treatment verification and beam-quality assessment. Microdosimetric characterization of therapeutic ion beams with solid-state detectors has been extensively investigated over the past decades, covering proton~\cite{Tran_2017_Proton}, helium~\cite{Petringa_2025_He}, carbon~\cite{Magrin_2020_diamond_measurements_MAUS}, and heavier ion beams~\cite{Lee_2021_Meas_CMRP}. However, routine clinical implementation requires a transition from application-specific research prototypes to dedicated readout systems designed to meet the constraints of realistic treatment environments. In particular, high-rate operation must be ensured to minimize pulse pileup, while maintaining sufficient spectroscopic resolution to characterize the full dynamic range encountered in a therapeutic setting.

\section{Experimental Microdosimetry}

Microdosimetry characterizes the quality of radiation fields by describing the stochastic distribution of energy deposition events in microscopic sites through lineal-energy spectra. The lineal energy, $y$, is defined as the energy imparted, $\Delta E$, by an individual radiation event within a target volume of specified size and composition, normalized by its mean chord length $\bar{\ell}_\text{path}$~\cite{ICRU_98_Microdosimetry}

\begin{equation}
y = \frac{\Delta E}{\bar{\ell}_\text{path}}.
\end{equation}

The frequency spectra $f(y)$ of lineal energy are typically obtained in a single-event mode from \gls{PHA} using charge-sensitive readout electronics. This constitutes a \gls{CSA}, which amplifies the generated signal charge into a voltage step proportional to the deposited energy in the \gls{SV}, followed by pulse shaping to optimize the frequency content for amplitude \gls{SNR} and subsequent digitization and pulse height extraction. Traditionally, different analog shaping circuits were used followed by a \gls{MCA}. However, modern high-speed, high-resolution \glspl{ADC} enable direct digitization of the preamplifier signal, followed by \gls{DSP} that provides more precise control over the shaping parameters and filtering. The histogram of the measured pulse heights directly yields the frequency lineal energy spectrum, $f(y)$, following proper calibration, which is most commonly performed using particle-edge calibration based on known spectral features~\cite{Conte_2013_Edge_Cal}. Normalized to area one, this corresponds to the probability distribution function for the energy deposition of a given particle in the radiation field. The dose distribution is derived from the frequency spectrum by weighting each event according to its lineal energy

\begin{equation}
d(y) = \frac{1}{\bar{y}_F} y f(y).
\end{equation}

Normalization by the frequency-mean lineal energy $\bar{y}_F$ ensures that the resulting distribution represents the relative contribution to the absorbed dose. For visualization, microdosimetric spectra are commonly presented using the weighted representations $yf(y)$ and $yd(y)$ in a semi-logarithmic plot~\cite{ICRU_98_Microdosimetry}. While gas-based \glspl{TEPC} used to be the reference detector for microdosimetry, their relatively large physical dimensions limit spatial resolution and increase pileup at the high particle fluxes typical of therapeutic ion beams (up to \SI{e10}{\per\second}).

\begin{figure}[htbp]
\centering
\includegraphics[width=0.8\textwidth]{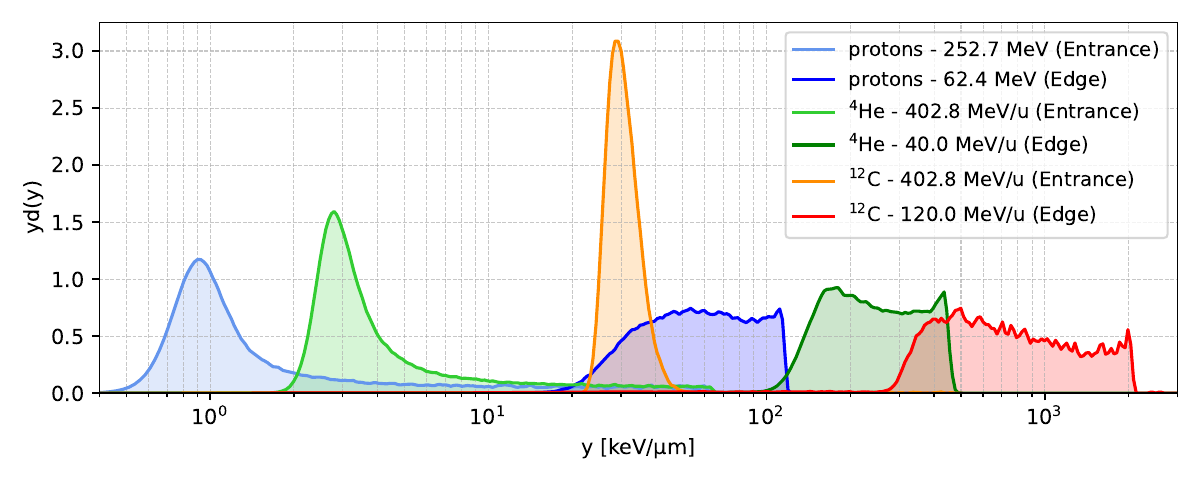}
\caption{The dynamic range of microdosimetric lineal-energy spectra in a typical light-ion therapy setting with protons, carbon-ions and helium-ions in a \SI{10}{\micro\meter} diamond detector. The entrance spectra at the highest energies and the particle edge~\cite{Conte_2013_Edge_Cal} spectra at the lowest available energies are shown for the example of MedAustron. The results are obtained from \gls{MC} simulations using Gate~\cite{GATE1}.\label{fig:dynamic_range}}
\end{figure}

Advances in microfabrication have established solid-state microdosimeters manufactured from silicon~\cite{Rozenfeld_2016_Overview_Wollongong_Si,Guardiola_2020_Overview_CNM_Si}, \gls{CVD}-diamond~\cite{Verona_2024_Diamond,Zahradnik_2020_LCD_Diamond,Davis_2017_Wollongong_Diamond}, and recently in \gls{SiC}~\cite{Petringa_2025_SiC_uDos,Venegas_2026_CNM_SiC_uDos}, whose true \si{\micro\meter}-scale sensitive volumes and rapid charge collection provide superior spatial resolution and reliable operation at clinically relevant dose rates. An overview is given in~\cite{Parisi_2022_Review_Microdosimetry}. However, despite significant technological advances, experimental microdosimetry in the context of light-ion beam therapy continues to face challenges due to the high dose rates and low \gls{SNR}.\\

\paragraph{Signal-to-Noise Ratio}
The signal generated in microdosimeters is inherently small given the dimensions of the \glspl{SV}. At the same time, clinical ion beams exhibit an exceptionally large dynamic range, extending from high-energy protons in the entrance channel and low-\gls{LET} contributions from delta-electrons to stopping carbon-ions near the Bragg peak. Figure \ref{fig:dynamic_range} shows examples for protons, helium-ions and carbon-ions at MedAustron. A \SI{250}{\mega\electronvolt} proton deposits on average only \SI{8.2}{\kilo\electronvolt} in a \SI{10}{\micro\meter}-thick diamond detector, corresponding to about \SI{625}{\electron}. To faithfully resolve low-\gls{LET} events, the readout electronics should therefore aim for an \gls{ENC} below \SI{100}{\electron}. Modern \glspl{CSA} in conjunction with optimized \gls{PCB} design and pulse shaping can meet these requirements.

\paragraph{Pileup}
The optimum shaping time for low-noise charge-sensitive readout is typically on the order of \SI{1}{\micro\second}, limiting the rate at which individual events can be resolved. Thus, obtaining pileup-free microdosimetric spectra at clinical dose rates remains challenging despite the small sensitive areas of modern microdosimeters, typically on the order of \SI{100}{\micro\meter}. Particle extraction from therapeutic accelerators follows a Poisson-like stochastic process, resulting in a finite probability of closely spaced events even at moderate average count rates~\cite{Knopf_2025_Rate_MAUS,Knopf_2026_Microspill}. Simply reducing the detector area is not a practical solution, as the acquisition time required to obtain sufficient counting statistics would become prohibitively long. Pileup can be mitigated using \gls{PUR} algorithms, the most common being the fast--slow coincidence method. In this approach, the preamplifier output is processed by two parallel shaping paths: a fast shaper with a shaping time on the order of \SI{50}{\nano\second} and a slow shaper with a shaping time of approximately \SI{1}{\micro\second}. The fast channel is used to detect closely spaced events and inhibits their acquisition, while the slow channel, optimized for maximum \gls{SNR}, is used to extract the pulse amplitude. An additional avenue for improving pileup rejection is offered by the intrinsic micro-spill time structure of extracted accelerator beams. The particle flux is modulated at the accelerator orbital frequency, which in synchrotrons is comparable to the optimum shaping time of the readout chain, allowing this temporal correlation to be exploited for improved event separation~\cite{Knopf_2026_Microspill}.

\section{The Spectacular Data Acquisition System}

Acquiring statistically meaningful spectra under clinically relevant conditions within practical timescales requires instrumentation capable of simultaneously handling high count rates, wide dynamic ranges, and stringent resolution requirements, ideally approaching \SI{0.1}{\kilo\electronvolt\per\micro\meter}. Satisfying these demands requires the combined optimization of the analog electronics and \gls{DSP}.\\

\paragraph{Hardware}

\begin{figure}[htbp]
\centering
\includegraphics[width=.45\textwidth]{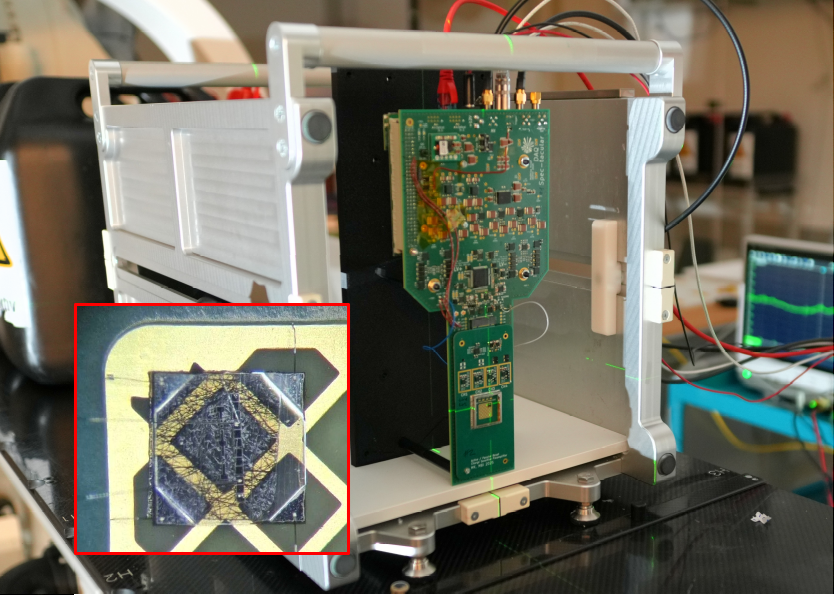}
\qquad
\includegraphics[width=.17\textwidth]{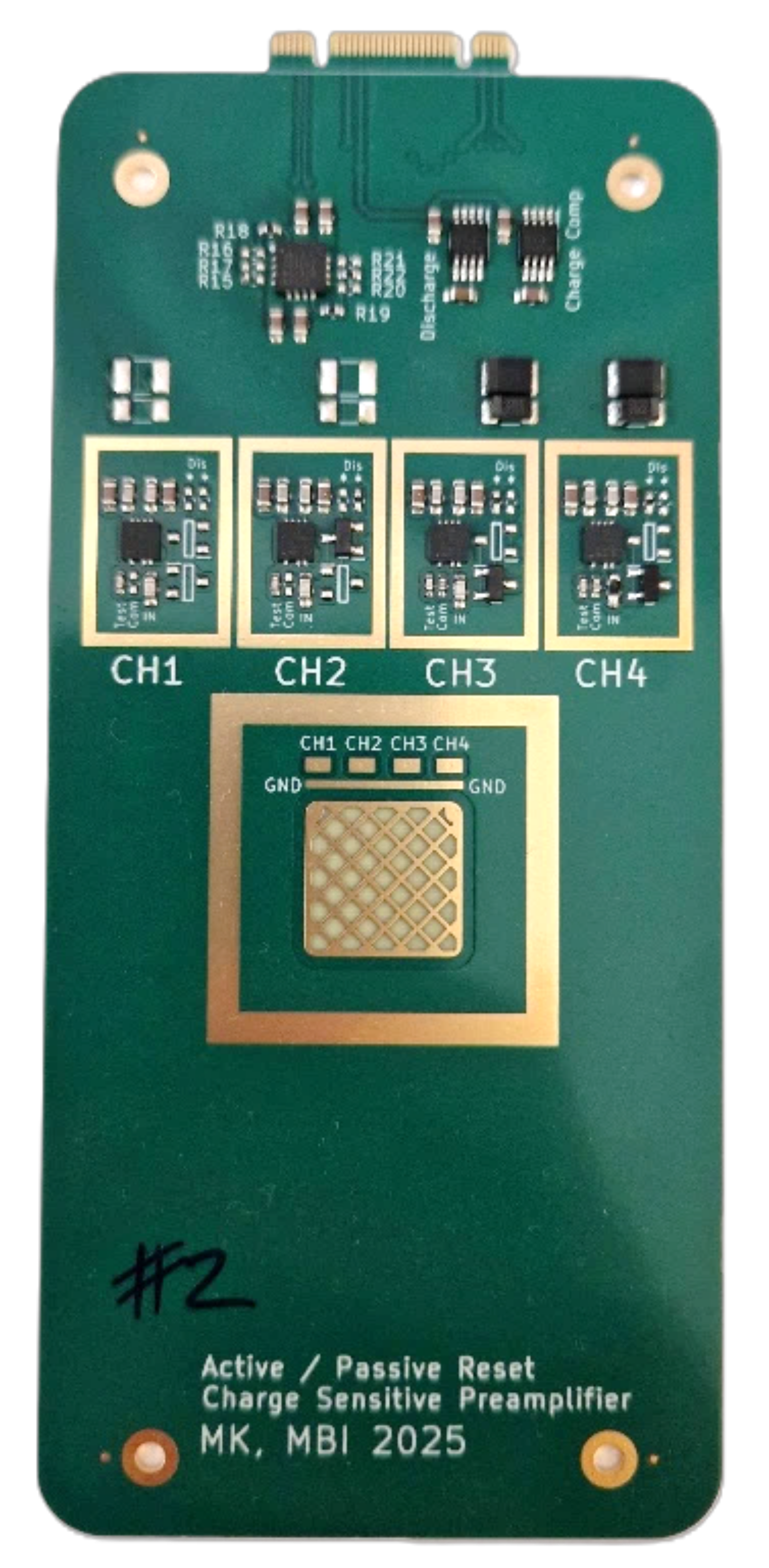}
\caption{Left: The prototype \emph{Spectacular} carrier board in the experimental room at MedAustron. The holder was used to put RW3 water-equivalent plates in front of the detector to achieve different radiation qualities. The inset shows the \SI{10}{\micro\meter} diamond microdosimeter array used for the presented measurements. Right: The analog front-end used for the presented measurements, featuring four custom \glspl{CSA} based around commercial op-amps.\label{fig:pcbs}}
\end{figure}

\emph{Spectacular} is based \color{black}around \color{black} a Xilinx Zynq UltraScale+ \gls{SoC}, enabling real-time signal processing and high-bandwidth streaming of transient signals digitized by a 16 bit \gls{ADC} (Analog Devices AD9446) operating at \SI{100}{\mega\sample\per\second}. A carrier board supplies power, control, and digitization capabilities for up to four analog front-end channels, which are hosted on separate daughter boards. These \color{black}daughter boards \color{black} can be hot-swapped during operation through a standardized board-edge connector (M.2 NGFF). The hardware was designed to support four analog channels. However, in this first prototype only one channel is digitized at a time and the input is multiplexed to the \gls{ADC} driver. \color{black}Simultaneous acquisition of multiple channels will be implemented in a future version as it is essential to enable a practical system. \color{black} To enable fast hardware development, the \gls{SoC} was integrated using a commercial development board (Trenz TE0706) to the back of the carrier board, which in turn hosts a \gls{SoM} (Trenz TE0820-05-4AE21MA). Four positive and four negative supply voltages, individually programmable up to $\pm$\SI{5}{\volt}, are provided to supply the analog front ends. The carrier board also features a test and a reset pulse generator capable of producing fast pulses with $\sim$ns rise time. Positive and negative pulses with amplitudes of up to $\pm$\SI{2.5}{\volt} can be routed to any of four dedicated front-end lines. These signals can be used either to inject test pulses into the front-end electronics or to provide an active reset for the \glspl{CSA}. In addition, several \gls{FPGA} I/O lines are reserved and are available for future extensions or custom functionality. The board further provides an analog monitoring output and a dedicated test input directly connected to the \gls{ADC} driver for the injection of test signals. The \glspl{PCB} can be seen in figure \ref{fig:pcbs}.

\paragraph{Software \& Gateware}

\begin{figure}[htbp]
\centering
\includegraphics[width=.55\textwidth]{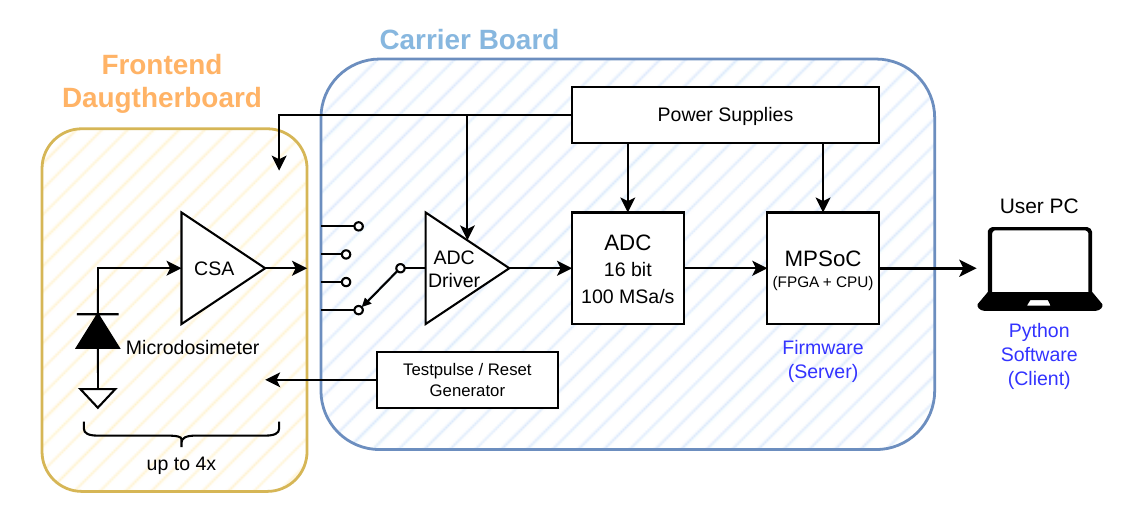}
\qquad
\includegraphics[width=.38\textwidth]{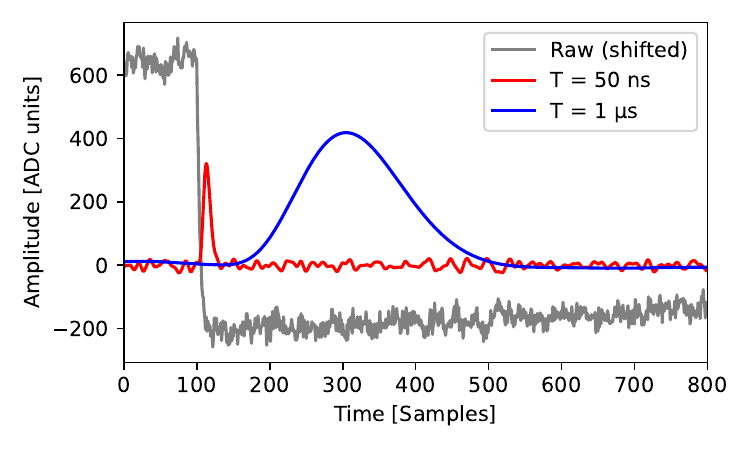}
\caption{Left: Architecture of the \emph{Spectacular} system. Right: Fast and slow quasi-Gaussian shaped pulses. \label{fig:arch_pulse}}
\end{figure}

The modular hardware architecture (figure \ref{fig:arch_pulse}) is complemented by a hierarchical software framework that provides a flexible processing environment in which algorithms can be deployed on the onboard \gls{FPGA}, the embedded CPU, or the host PC depending on the required performance and latency. To accommodate the continuous data stream of \SI{200}{\mega\byte\per\second}, the data are delta-compressed on the \gls{FPGA} before being packaged in a custom format and provided via \gls{DMA} to the CPU. The firmware running on the \gls{SoC} then streams the data over a Gigabit Ethernet socket to the host PC running the Python control software. This user software also provides functionality for slow control and live visualization and evaluation of the digitized data stream in a modular GUI. Furthermore, all acquired data can be permanently stored in a compressed custom binary format. Acquisitions can either be performed in an untriggered mode or synchronized with the accelerator extraction trigger signal. The \gls{FPGA} also controls the active reset of the \glspl{CSA} by detecting threshold crossings in the digitized signal.\\

\paragraph{Analog Front-End}

The long-term goal is to integrate a microdosimetric \gls{ASIC} with \color{black}the \emph{Spectacular} system\color{black}~\cite{Waid_ASIC}. Given the extended development cycle, a discrete \gls{CSA} based on commercially available components was developed for initial testing. It is based on a low-noise FET-input operational amplifier (Analog Devices OPA817), configured as a \gls{CSA} with a small feedback capacitance $C_f=$\SI{50}{\femto\farad} and a passive reset using a $R_f=$\SI{1}{\giga\ohm} feedback resistor. To avoid \gls{ADC} saturation during sustained pulse trains, an additional active reset option was implemented using pico-ampere diodes connected to the input node, enabling \gls{FPGA}-controlled discharge of the feedback capacitance. The \gls{ENC} of the \gls{CSA} was determined with the sensor attached and installed in the irradiation room at MedAustron from the \gls{FWHM} of the baseline fluctuations. At the experimentally established optimal shaping time of \SI{1}{\micro\second}, the amplifier achieves an \gls{ENC} of $\sim$\SI{175}{\electron}. \color{black} This performance is not yet sufficient to fully resolve low-\gls{LET} events, and further reduction of the electronic noise will therefore be a key objective for the next hardware iteration. \color{black} The capacitance of the sensor amplifier interconnect is expected to contribute significantly to this value and has been reduced in an upcoming \gls{PCB} revision. Linearity of the whole readout chain was verified by injecting calibrated charge pulses through a \SI{0.1}{\pico\farad} test capacitor at the input node. No saturation was observed up to an injected charge of \SI{250}{\femto\coulomb}, roughly corresponding to the highest possible energy deposition of a carbon-ion in a \SI{10}{\micro\meter}-thick diamond \gls{SV} (carbon edge)~\cite{Conte_2013_Edge_Cal}.

\paragraph{Processing}

For the presented measurements, a fourth-order quasi-Gaussian shaper is synthesized following the approach described in~\cite{Ohkawa_1976}. From the resulting transfer function, a \gls{FIR} filter kernel is constructed to enable fast online filtering of the preamplifier signal. The amplitudes were extracted using a \SI{1}{\micro\second} shaping time. \Gls{PUR} was implemented using a \SI{50}{\nano\second} fast shaping function and inhibiting the acquisition of subsequent pulses up to the fourth order (see figure \ref{fig:arch_pulse}).

\section{Characterization at the MedAustron Ion Therapy Center}

\begin{figure}[htbp]
\centering
\includegraphics[width=.45\textwidth]{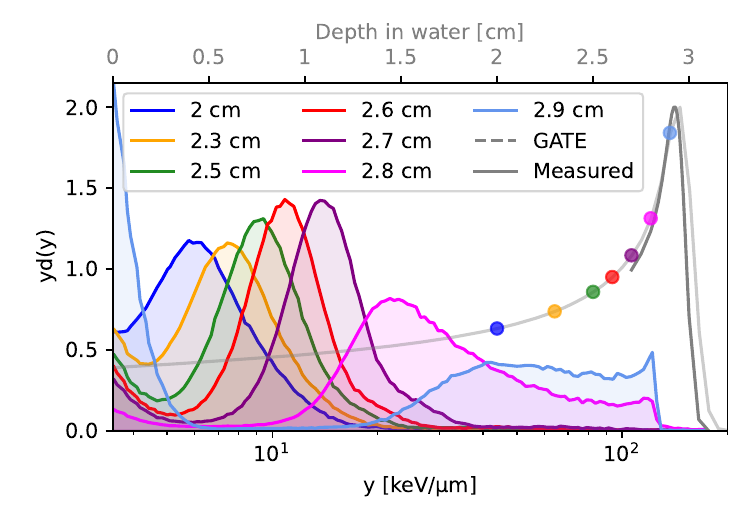}
\qquad
\includegraphics[width=.45\textwidth]{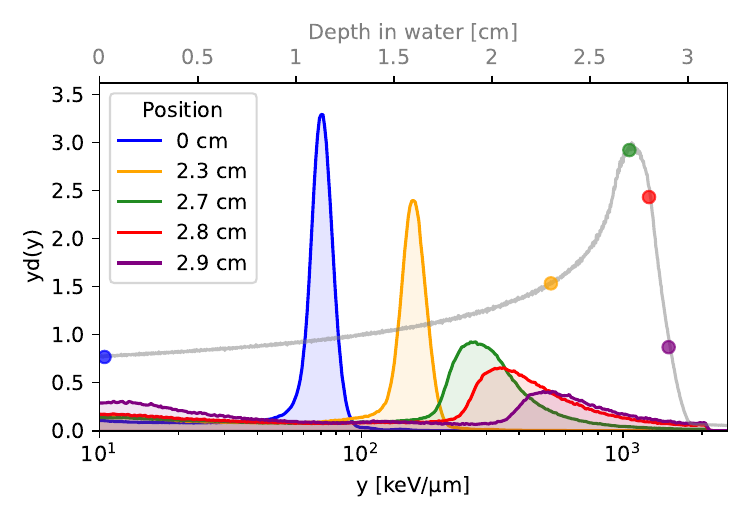}
\caption{Microdosimetric spectra along the depth-dose curve in water-equivalent RW3 plates acquired for \SI{62.4}{\mega\electronvolt} proton (left) and \SI{120}{\mega\electronvolt\per\u} \ce{^12C^6+}-ion (right) beams at MedAustron. The position is indicated on the depth-dose curves obtained from \gls{MC} simulations using Gate 9.3~\cite{GATE1}. \label{fig:results}}
\end{figure}

A \SI{10}{\micro\meter}-thick diamond microdosimeter was used for the measurements presented in this work. The detector was fabricated at the Department for Industrial Engineering of the University of Tor Vergata (Rome) using a \gls{CVD} process~\cite{Verona_2024_Diamond}. It comprises an array of twelve independent \qtyproduct{160 x 160}{\micro\metre} Schottky diodes defined by chromium top electrodes. The common back electrode consists of a highly boron-doped diamond layer. During the measurements, one diode was reverse-biased at \SI{15}{\volt} using a Keithley 2470 SMU and DC-coupled to the input of the \gls{CSA}. Microdosimetric spectra were acquired for \SI{62.4}{\mega\electronvolt} protons and for \SI{120}{\mega\electronvolt\per\u} \ce{^12C^6+}-ion beams using a single-spot beam at the iso-center under full clinical dose-rate conditions. Different radiation qualities were obtained along the depth-dose curve by placing water-equivalent RW3 "solid water" plates (PTW Freiburg, Germany) in front of the detector (see figure \ref{fig:pcbs}). The spectra were calibrated using the proton edge~\cite{Conte_2013_Edge_Cal}, which was well resolved in the spectrum acquired at a \gls{WET} of \SI{2.9}{\centi\meter}. For carbon-ions, the full dynamic range could be covered. For protons, however, the noise performance is yet insufficient to fully resolve low-\gls{LET} particles in the entrance channel, and a conservative lower cutoff of \SI{3.5}{\kilo\electronvolt\per\micro\meter} was applied. Consequently, only spectra acquired after \SI{2}{\centi\meter} \gls{WET} could be fully resolved. With ongoing revisions of both the analog front-end and carrier board, together with the integration of a chip-based \gls{CSA}, this previously inaccessible range is expected to be uncovered. The results are shown in figure \ref{fig:results}.

\section{Conclusion and Outlook}

Despite the distinct requirements of microdosimetry, dedicated readout electronics remain scarce. The MicroPlus probe~\cite{Rozenfeld_2016_Overview_Wollongong_Si} and the DIODE detector~\cite{Verona_2025_DIODE} represent notable exceptions. \color{black}Nevertheless, both systems are limited to integrating the analog front-end. Digitization and further signal processing are delegated to external hardware. \color{black} The realization of a complete, modular, and easily extendable hardware platform represents an important advancement in the field. The long-term objective is the integration of the complete analog chain into a dedicated \gls{ASIC}, enabling improved performance through a tighter control of the parameters. \color{black} As an initial implementation, the first prototype of the \emph{Spectacular} system has demonstrated the feasibility of performing microdosimetric measurements with different ion beams at therapeutic dose rates, with promising energy resolution. While the present prototype still has limitations in electronic noise and simultaneous multi-channel acquisition, further hardware iterations, including a revised carrier board and multiple analog front-end designs, are currently under development to improve the performance of the system. \color{black} In particular, the integration of a dedicated \gls{CSA} \gls{ASIC}~\cite{Waid_ASIC}, which is under characterization at the time of writing, is anticipated to further improve the achievable signal resolution. In parallel, the software framework is continuously being expanded to simplify data acquisition, speed up processing, and facilitate operation by non-expert users. Future developments will also focus on migrating a larger fraction of the \gls{DSP} to the \gls{FPGA}, providing the scalability required for simultaneous multi-channel readout and live processing. \color{black}A range of novel \gls{SiC} and diamond microdosimeters are planned to be integrated with \emph{Spectacular}. \color{black} Ultimately, this work aims to contribute towards making microdosimetric measurements a more accessible and practical tool for generating experimental data relevant to clinical ion-beam applications.

\acknowledgments
The financial support of the Austrian Ministry of Education, Science and Research is gratefully acknowledged for providing beam time and research infrastructure at MedAustron. The authors acknowledge TU Wien Bibliothek for financial support through its Open Access Funding Programme. This project has received funding from the Austrian Research Promotion Agency FFG, Austria, grant number 918092.

% Bibliography
\bibliographystyle{JHEP}
\bibliography{biblio.bib}

\end{document}